\documentclass[aps,twocolumn]{revtex4-1}
\usepackage{epsfig}
\usepackage{color}
\usepackage{amsmath}
\usepackage[retainorgcmds]{IEEEtrantools}

\begin{document}

\title{Partial inertia induces additional phase transition
in the explosive majority vote model}

\author{Pedro E. Harunari$^1$, M. M. de Oliveira$^2$ and 
C. E. Fiore$^1$}
\email{fiore@if.usp.br} 

\affiliation{$^1$ Instituto de F\'{\i}sica,
Universidade de S\~{a}o Paulo, \\
Caixa Postal 66318\\
05315-970 S\~{a}o Paulo, S\~{a}o Paulo, Brazil\\
$^2$Departamento de F\'{\i}sica e Matem\'atica,
CAP, Universidade Federal de S\~ao Jo\~ao del Rei,
Ouro Branco-MG, 36420-000 Brazil. }
\date{\today}

\begin{abstract}
Recently it has been aroused a great interest about
explosive (i.e., discontinuous)  transitions.
They manifest in distinct systems, such as
synchronization in coupled oscillators, percolation regime, absorbing
phase transitions and
more recently, in the majority-vote (MV) model
with inertia.  In the latter, the model rules are slightly modified 
by the inclusion
of a term depending on the local spin (an {\it inertial term}).
 In such case, Chen et al. (Phys Rev. E {\bf 95}, 042304 (2017))
have  
found  that relevant inertia changes the nature of the phase
transition in complex networks, from continuous to discontinuous.   
Here we  give a further step  by  embedding inertia  only in vertices 
with degree larger than a threshold value $\langle k \rangle k^*$, 
$\langle k \rangle$ being the mean system degree and $k^*$ the fraction restriction.
 Our results, from mean-field  analysis and extensive
numerical simulations, reveal that an explosive 
transition is presented in both  homogeneous and heterogeneous
structures for small and intermediate $k^*$'s. Otherwise,
 large restriction can sustain a
 discontinuous transition only in the heterogeneous case. This
 shares some similarity with recent results for the Kuramoto model
(Phys Rev. E {\bf 91}, 022818 (2015)).
Surprisingly, intermediate restriction and large inertia are responsible
for the emergence of  an extra phase, in which the system is
partially synchronized and the
classification of phase transition depends on the
inertia  and the lattice topology.
In this case, the system exhibits two phase transitions.
\end{abstract}

\maketitle

\section{Introduction}
An explosive (discontinuous)   transition 
occurs when an infinitesimal increase of the
control parameter produces an abrupt change  in macroscopic
quantities. This kind of transition has attracted 
a lot of interest in the recent years, inspired by
the discovery of a procedure (the ``Achlioptas process") that gives rise to an abrupt percolation transition in complex networks \cite{explosive1,explosive2,explosive3,explosive4}.
While subsequent works have shown the Achlioptas process transition was, in fact, a continuous phase transition with unusual finite-size scaling \cite{continuous1,continuous2,continuous3}, many related models with alternative mechanisms showing genuinely discontinuous and anomalous transitions have now been discovered (see Ref. \cite{expreview} and references therein).

One of these main examples appear in the context of coupled 
oscillators in which  the Kuramoto Model (KM)
\cite{kuramoto} plays a central role. 
 The original KM describes self-sustained coupled phase
oscillators and exhibits a continuous phase transition at a critical
coupling, beyond which a collective behavior is achieved. 
A few years ago, in a pioneering work, Garde\~nes et al. \cite{prl2011},
discovered
 that a discontinuous phase transition to synchronization emerges as a consequence of the correlation 
 between structure and local dynamics when a scale-free network is considered.
Subsequent studies have confirmed the transition robustness
under changing ingredients, such as lattice topology \cite{prl2011},  time delay \cite{peron}, disorder 
\cite{skardal} and inertia \cite{inertia}. Analysis of the explosive transition in simpler structures, such
as star graphs, for which exact treatment is possible \cite{tiago}, also confirmed that  the transition to collective behavior is  discontinuous. 
Investigating the explosive synchronization in a generic complex network, Zhang et al. \cite{kurths} have found that a positive correlation between  
the oscillators frequency and the degree of
their corresponding vertices is the required condition 
for its appearance.
More recently, Pinto et al. \cite{saa} have verified that
it suffices to fulfill above minimal requirement for
the hubs (e.g. the vertices with  higher degrees)
for promoting an abrupt transition.

Besides dynamical systems, they
manifest in  markovian nonequilibrium
reaction-diffusion processes. Two groups are important
in this context: those presenting absorbing states and 
the up-down symmetric systems. In the former, distinct mechanisms, such as the inclusion of a quadratic term in the particle 
creation rates \cite{quadratic1,quadratic2},  the need of a minimal
neighborhood for generating subsequent offsprings \cite{fiore14}, synergetic effects in multi species models \cite{scp1,scp2} or cooperative coinfection in multiple diseases epidemic models \cite{coinfect1,coinfect2,coinfect3} 
can be taken into account for shifting, 
from a  continuous transition (belonging generically to the directed
percolation (DP) universality class \cite{marr99,odor07,henkel})
to a discontinuous one.

The Majority Vote (MV) model is one of
the simplest nonequilibrium up-down symmetric systems exhibiting 
an order-disorder phase transition \cite{mario92}. 
Extensive studies of this model in distinct lattice topologies 
(besides the usual regular ones)
showed that the symmetry-breaking phase transition
is not affected by the kind of the underlying networks \cite{chen1},
although the critical behavior results in  
set of critical exponents entirely different \cite{pereira}.
However, very recently, Chen et al. \cite{chen2} verified that the usual
second-order phase transition in the majority vote (MV) 
becomes first-order when 
a term depending on the local density is included in the dynamics
(an {\em inertial} effect).

Aimed at investigating how the network topology and inertial effects contribute
 to the emergence of the explosive transition in the MV model, 
in this work we include the inertia only in a given fraction of sites
with degree $k$ larger than a threshold $\langle k \rangle k^*$.
We observe that  the MV  transition remains 
explosive only for a low/intermediate fraction
of restriction $k^*$ in homogeneous
structures.  On the other hand, in heterogeneous networks, it is 
sufficient to include inertia only in the hubs for promoting an abrupt 
behavior.
Remarkably, a new feature  induced by the partial (but large) 
inertia is the emergence
of an extra phase, in which the system is partially synchronized, whose
phase transition can be continuous or discontinuous, according to the
inertia magnitude. In this region
of the phase diagram, the system presents two phase transitions.

This paper is organized as follows: In Sec. II, we derive
the mean field theory for the model. Next, numerical results are shown in Sec. III. 
Conclusions are drawn in Sec. IV.

\section{Model and mean field analysis}

The  MV model is defined in
an arbitrary lattice  topology, in which
each node $i$ of degree $k_i$ is attached to a
spin variable, $\sigma_i$,  that can take the values 
$\sigma_i=\pm 1$. In the original case, with probability $1-f$
each node $i$ tends  to align itself with its local neighborhood majority,
 and with complementary probability $f$, the majority rule is not
followed.  The quantity $f$ is a misalignment term whose increasing
gives rise to an order-disorder (continuous) phase transition.
Chen et al.  \cite{chen2} added  to the transition rate a term
 depending only on the
local state $\sigma_i$, {\em irrespectively}
the majority nearest neighbor spins.
Mathematically, one has the following transition rate
\begin{equation}
  w(\sigma_i)=\frac{1}{2}\left\{1-(1-2f)\sigma_{i}
  S\left[(1-\theta)\sum_{j=1}^{k_i}\sigma_j/k_i+\theta \sigma_i\right]\right\},
\label{eq2}
\end{equation}
where $\theta$ denotes the inertia strength and 
$S(X)$ is defined by $S(X)={\rm sign(X)}$  
 if $X \neq 0$ and $S(0)=0$.
Note that for $\theta=0$ one recovers the original MV model,
whose critical transition  depends on the nodes distribution.
In particular, for $\theta > 0.5$, 
the dynamics is fully dominated by the inertia and no 
phase transition is observed \cite{chen2}. 

The time evolution of  the  density $\rho_k$ of
``up'' spins ($\sigma_i= 1$) of a  node with degree $k$ given by 
\begin{equation}
\frac{d }{dt}\rho_k=w_{-1 \rightarrow 1}(1-\rho_k)-\rho_k w_{1 \rightarrow -1},
\label{eq3}
\end{equation}
where $w_{-1 \rightarrow 1}$ and $w_{1 \rightarrow -1}$ denote the transition
rates to  states with opposite spin. In the steady state one has that
\begin{equation}
\rho_k=\frac{w_{-1 \rightarrow 1}}{w_{-1 \rightarrow 1}+
w_{1 \rightarrow -1}}.
\label{eq4}
\end{equation}

The average magnetization of a node of degree $k$  
is related to $\rho_k$ through the relation
$m_k=2\rho_k-1$.
Our first inspection of the inertia effect is carried out
through a mean-field treatment.
Here, we follow the ideas from Ref. \cite{romualdo,chen2},  in which
the transition rates in Eq. (\ref{eq3}) are rewritten in terms of
the majority and minority rules, given by
\begin{equation}
w_{-1 \rightarrow 1}=(1-2f){\bar P_{-}}+f,
\end{equation}
where ${\bar P_{-}}$  is the probability that the node $i$ of degree $k$, with spin
$\sigma_i=-1$, changes its state according to the majority rule. In
particular, ${\bar P_{-}}$ depends  on the number $n_{k}^{-}$
of nearest neighbor $+$ spins 
in such a way that
${\bar P_{-}}=\sum_{n=\lceil n_k^{-} \rceil}^{k}(1-\frac{1}{2}\delta_{n,n_{k}^{-}})C_{n}^{k}
p_{+1}^{n}p_{-1}^{k-n}$, where $p_{\pm 1}$ is
the probability that one of its nearest-neighbors is in 
the spin state $\pm 1$.  Since the quantity
$n_{k}^{-}$ corresponds to the lower limit, it is evaluated
from  the condition $S(X)=0$, reading $n_{k}^{-}=\frac{k}{2(1-\theta)}$.

A similar expression  is obtained by writing down
the transition rate $w_{1 \rightarrow -1}$ in terms of the minority rule
(instead of the majority one)
\begin{equation}
w_{1 \rightarrow -1}=(1-f)-(1-2f){\bar P_{+}},
\end{equation}
where  ${\bar P_+}$ also reads 
${\bar P_{+}}=\sum_{n=\lceil n_k^{+} \rceil}^{k}(1-\frac{1}{2}\delta_{n,n_{k}^{+}})C_{n}^{k}
p_{+1}^{n}p_{-1}^{k-n}$, but the bottom limit reads
$n_{k}^{+}=\frac{k(1-2\theta)}{2(1-\theta)}$.

For large $k$, each term of the  above binomial distributions 
 ${\bar P_{\pm}}$ approach to  gaussian ones
with mean $m=kp_{+}$ and  variance $\sigma^{2}=kp_{+}p_{-}$. 
So that
\begin{equation}
{\bar P_{\pm}} \rightarrow \frac{1}{2}\left\{1 \pm {\rm erf}\left[
\sqrt {2k}\left(\frac{\theta}{2(1-\theta)} \pm y \right) \right]\right\},
\label{eq7}
\end{equation}
 where ${\rm erf(x)}$ denotes the error function and  
the nearest neighbor probability $p_+$ has been rewritten in terms of the
quantity $y$ through the relation $p_+\equiv 1/2+y$.

For any node without degree correlation, the probability
that a randomly nearest neighbor has degree $k$
is $kP(k)/\langle k \rangle$. Thus, $p_+$ and $\rho_k$ 
are related by $p_+=\sum_{k}kP(k)/\langle k \rangle \rho_k$
and finally we arrive at the following expression
\begin{equation}
y+1/2=\sum_{k}\frac{k P(k)}{\langle k \rangle}
\frac{(1-2f) {\bar P_-}+f}{1+(1-2f)({\bar P_-}-{\bar P_+})},
\label{eq8}
\end{equation}
with ${\bar P_\pm}$  evaluated from Eq. (\ref{eq7}). By splitting
  Eq. (\ref{eq8}) in two parts, the first and
  second terms get restricted to the nodes {\em in the absence of} and {\em with} inertia
respectively, in such a way that
\begin{widetext}
\begin{equation}
  y+1/2=\sum_{k=k_{0}}^{\langle k \rangle k^{*}-1}
\frac{kP(k)}{2\langle k \rangle}[1+(1-2f){\rm erf}(\sqrt{2k}y)]+
  \sum_{k=\langle k \rangle k^{*}}^{\infty}
\frac{kP(k)}{\langle k \rangle}\frac{1-(1-2f){\rm erf}\left[
\sqrt {2k}\left(\frac{\theta}{2(1-\theta)} - y \right) \right]}
{2-(1-2f)\{{\rm erf}\left[
\sqrt {2k}\left(\frac{\theta}{2(1-\theta)} - y \right) \right]+{\rm erf}\left[
\sqrt {2k}\left(\frac{\theta}{2(1-\theta)} + y \right) \right]\}},
\label{eq9}
\end{equation}
\end{widetext}
where $k_0$  denotes the minimum degree. 
In particular,
for $\theta=0$, Eq. (\ref{eq9}) reduces to $y=\frac{1-2f}{2\langle k \rangle}
\left\{\sum_{k}k[{\rm erf}(\sqrt{2k}y)]P(k)\right\}$,
in consistency with results from Refs. \cite{chen1,chen2}.

Thus,  the solution(s) of Eq. (\ref{eq9}) give
us the values of $y$, whose  corresponding $\rho_k$'s 
are obtained from Eq. (\ref{eq4}).
The mean magnetization $|m|$ is achieved  by
 summing over all values of $k$ 
 with their correspondent weights $P(k)$, so
 that $|m|=2\left[\sum_{k}\rho_kP(k)\right]-1$. 

\begin{figure}[h!]
\epsfig{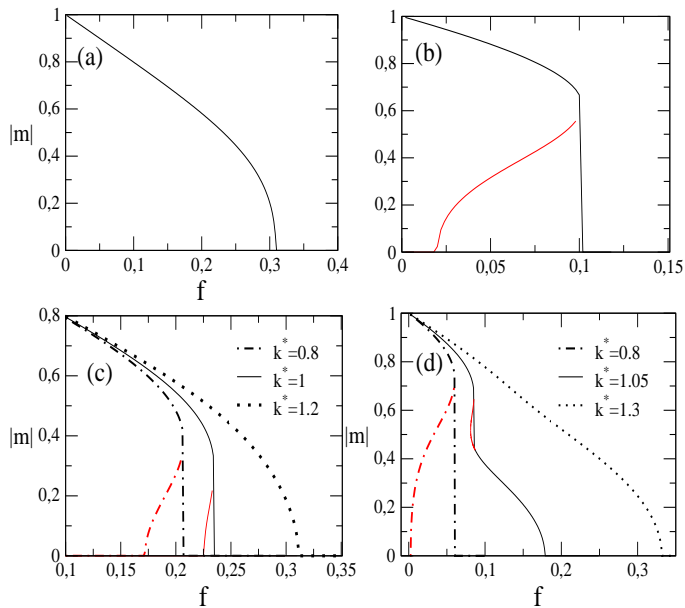}
\caption{For the ER and $\langle k\rangle=20$,
  $|m|$ versus $f$ for distinct inertia rates. 
  Panels $(a)$ and $(b)$ show
  the full inertia cases for $\theta=0.2$ and $0.4$, respectively.
  In panels $(c)$ and $(d)$, the restrictive case
   for $\theta=0.3$ and
   $\theta=0.45$ and distinct restrictions $k^*$'s. In all cases,
red lines correspond the unstable branches.}
\label{fig1}
\end{figure}

\begin{figure}[h!]
  \epsfig{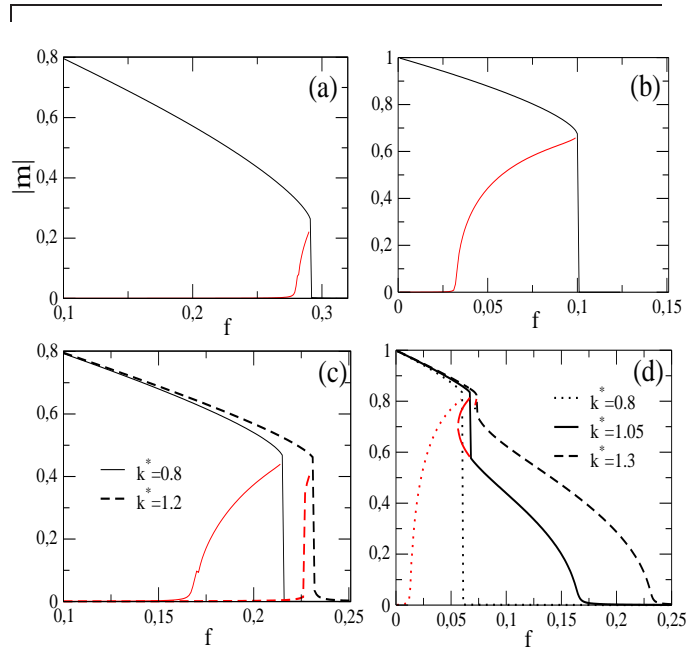}
\caption{For the PL and $\langle k\rangle=20$,
  $|m|$ versus $f$ for distinct inertia rates. 
  Panels $(a)$ and $(b)$ show
  the full inertia cases for $\theta=0.2$ and $0.4$, respectively.
  In panels $(c)$ and $(d)$, the restrictive case
   for $\theta=0.3$ and
   $\theta=0.45$ and distinct restrictions $k^*$'s. In all cases,
red lines correspond the unstable branches.}
\label{fig1-2}
\end{figure}

Figs. \ref{fig1} and \ref{fig1-2} show (for  $\langle k\rangle=20$), the behavior
of $|m|$ versus $f$, for two distinct network topologies.
The first is an Erdos-Renyi (ER) graph, a prototypical model of a homogeneous random network, 
with the degree distribution given by
$P(k)=\langle k \rangle^{k}e^{-\langle k \rangle}/k!$.
 The second case is  a 
representative description of heterogeneous networks, in which
nodes are distributed according to the probability distribution 
$P(k) \sim k^{-\gamma}$. From now on, such case will be referred 
as power law (PL) graph. Here, we take $k_0=0$ for the ER and,
 for avoiding divergences when $k \rightarrow 0$ in the PL, 
we have imposed  $k_0$  constrained to the mean 
degree $\langle k \rangle$ through the relation
$k_0=\frac{\gamma-2}{\gamma-1}\langle k \rangle$. 

As for ER and PL cases, top panels $(a)-(b)$  correspond to the full
inertia cases whose phase transition is continuous
 for low $\theta$ and its increasing gives rise to
a discontinuous one. Its emergence is signed by 
the appearance of
an unstable branch (red  lines),
ending at  to lower $f$'s when $\theta$ goes up. 
Despite the similarity between both cases, note
the transition and the crossover (from continuous to discontinuous)
points depend on $P(k)$.
For example, for $\theta=0.2$ the transition is continuous
for the ER and discontinuous  for the PL.
All these results are  consistent with those 
obtained in Ref. \cite{chen2}. 

Next, we examine the partial inertia case,
in which the inertia appears only in a specific 
fraction of nodes (the ones with  larger degrees).
This analysis is inspired by the work by Pinto et al. \cite{saa}
for the KM, in which
a positive correlation between frequency-degree  taken only for the hubs
is enough for promoting an explosive transition.
 For instance, we introduce 
the fraction parameter $k^{*}$, in such a way that 
$\theta \neq 0$ only if $k \ge  \langle k \rangle k^{*}$
and $\theta=0$ otherwise.  Extremely large  $k^{*}$ implies that most of the
nodes will be absent of inertia and thus the phase transition is 
expected to be similar to the $\theta=0$ case (continuous). 
In the opposite case, low $k^{*}$ makes the majority
of sites to have inertia and one expects a scenario around the
panels $(a)$ and $(b)$.
 Panels $(c)$ and $(d)$ in Figs. \ref{fig1} and \ref{fig1-2}
show the results of the mean field theory (MFT) for 
the ER and PL [$\langle k\rangle=20$] and distinct  sorts 
of $k^{*}$.

As expected,   for low  $k^{*}$'s  MFT predicts similar  behaviors 
than the full inertia cases (see, e.g., panels $(c)$ and $(d)$
for $k^*=0.8$).
However, the increase of restriction leads to opposite scenarios.
Whenever the discontinuous phase transition is suppressed for the ER
when $k^*=1.2$, it is maintained for the PL. Despite a similar  fraction of
nodes with inertia (about $21\%$ and $17\%$
for the ER and PL, respectively), the
presence of hubs in the PL
 sustains the  discontinuous transition.

Surprising,  an additional   phase transition
emerges for  intermediate $k^{*}$'s and large $\theta$. 
This is clearly exemplified for $\theta=0.45$ and $k^{*}=1.05$ [panels $(d)$] in 
which the presence
of a  jump and an unstable envelope signals a discontinuous 
transition between two synchronized phases for low $f$
($|m| \neq 0$ in both cases) followed by a smooth vanishing of
$|m|$ for large $f$.  In all cases, the transition between partially-ordered 
and disordered phase is continuous.

\begin{figure}[h!]
\epsfig{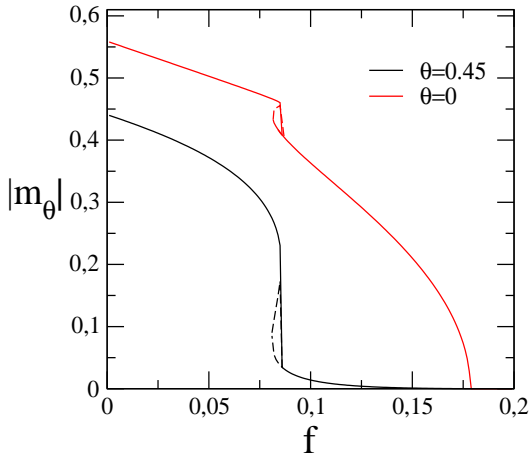}
\caption{For the ER, $\langle k\rangle=20$ 
and $k^*=1.05$, the plot of the 
right side of Eq. (\ref{eq9}) versus $f$ restricted to the inertial
part ($\theta=0.45$ in the present case) and $\theta=0$. 
Dashed lines correspond to the unstable solutions.}
\label{fig1-5}
\end{figure}

 To examine such a new feature in more details, we plot in 
Fig. \ref{fig1-5} the contribution of each part 
on  the right side of Eq. (\ref{eq9}) separately. 
That reveals  the former transition comes from
the subsystem with inertia (which becomes disordered), 
whereas the   other  gets ordered. By keeping the increase
of $f$, the  remaining  subset (without inertia) also loses  the ordering
at $f_c$.
In the next section, we continue the study of the effects 
of partial inertia through numerical simulations, 
in order to compare with our MFT predictions.

\section{Numerical results}

We performed extensive numerical simulations of the MV model on random graphs 
with ER and  PL topologies 
and system sizes $N$ ranging from $N=3600$ to $20000$.
 To construct an ER graph, we connect each pair
 of nodes with probability $\langle k \rangle/N$.
 When the size of the graph tends to infinity $N\to\infty$, 
its degree distribution is Poissonian, with mean $\langle k \rangle$. 
The PL graphs were generated using the uncorrelated 
configuration model (UCM) \cite{ucm}, so the degrees are
uncorrelated.  As in the MFT, we use $\gamma=3$ and distinct sets of inertia values.

 In order to classify the phase transition,  we begin by analyzing the absolute value of the mean
  magnetization per site $|m|$ as a function of $f$,  starting from a 
full ordered phase ($|m|= 1$) and increasing $f$ towards the completely
  disordered phase ($|m|=0$).
  In sequence, we take the opposite case,
wherein the system is in the disordered phase and the
parameter $f$ is gradually decreased approaching to the  ordered phase.
 Both increasing (``forward'') and decreasing (``backward'')
 curves are expected to coincide 
 when the phase transition is critical, but
 they are different at the phase coexistence (a trademark of a
 discontinuous transition). The presence of hysteresis
 indicates the system bistability with respect to the 
ordered/disordered phase according to its initial condition.
 They are also signed by the presence of 
 a bimodal probability distribution of the order-parameter $P(m)$. 
On the other hand, if the phase transition is continuous, $P(m)$
 exhibits a single peak, whose position depends on $f$.
 
Another feature distinguishing them relies  that
 the continuous case  presents an algebraic divergence
 of its order parameter variance 
$\chi=N[\langle m^{2} \rangle-\langle  m \rangle^{2}]$
 at the critical point $f_c$ \cite{martins-fiore}.  
(In simulations of finite systems, we observe a maximum 
that increases with the system size $N$).
The transition point $f_c$ can also be identified
through the reduced cumulant
$U_4=1-\frac{\langle m^{4} \rangle}{3\langle m^{2} \rangle^{2}}$,
since curves for distinct  $N$'s cross at   $f_c$. Off the critical
point, $U_4 \rightarrow 2/3$ and $0$
for the ordered and disordered phases, respectively when $N \rightarrow\infty$.

\begin{figure}
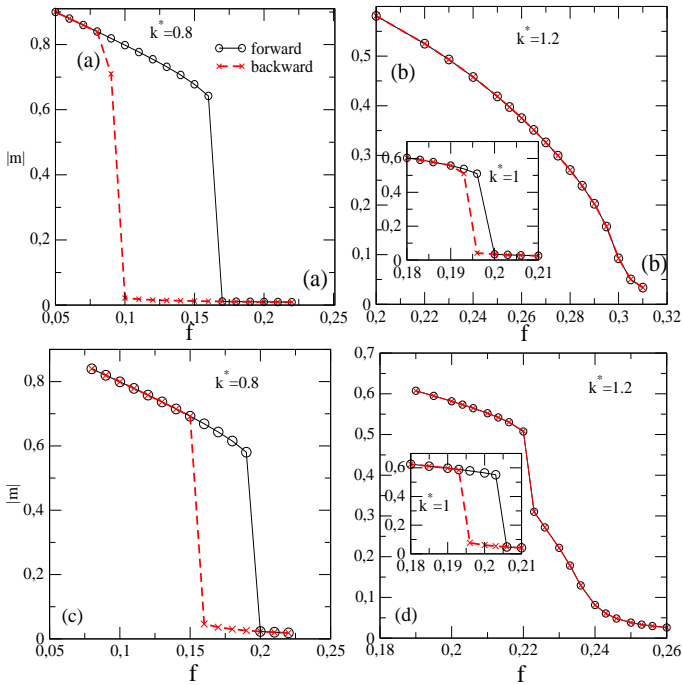

\epsfig{file=fig3-1.eps,width=9.cm,height=4.5cm}
\epsfig{file=fig3-2.eps,width=9.cm,height=4.5cm}
\caption{For ER [$(a)-(b)$] and PL [$(c)-(d)$] 
networks with $\theta=0.3$, the order parameter 
  $|m|$ versus the control parameter $f$
  for distinct $k^*$.  Circles  (stars) correspond the increase 
(decrease) of $f$
starting from an(a) ordered (disordered) phase.
  Except in panel $(d)$ ($N=40000$), results are  for $N=20000$.}
\label{fig2}
\end{figure}

In order to compare with the MFT, Fig. \ref{fig2}  shows results for
the ER and PL topologies and lower inertia values
 (exemplified here for $\theta=0.3$).
In both cases, the presence of hysteresis for  
$k^*=0.8$ and $k^*=1$ reveal, in similarity
with MFT,  discontinuous transitions
with hysteretic loop decreasing by 
elevating $k^*$.
Also,  the network
structure leads to opposite features for   $k^*=1.2$ with
a continuous phase transition for the ER (panel $(b)$).
The behavior  is  different for
the PL (panel $(d)$), showing a  small 
jump of $|m|$ at $f \sim 0.22$ for a partially ordered phase 
(see panel (d) in Fig. \ref{fig2}). This also contrasts with MFT results,
in which a discontinuous order-disordered transition is predicted.
Despite the evidence of a discontinuous transition for the PL,
we believe that sufficient larger $N$'s are
required for observing a hysteretic loop for such case.

\begin{figure}[h!]
\epsfig{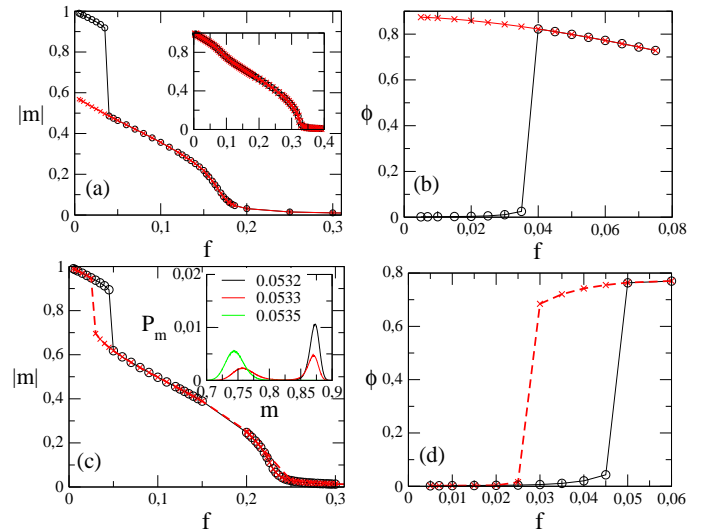}
\caption{For the ER network of size $N=20000$ and $\theta=0.45$, 
panels $(a)$ and $(b)$ show  
$|m|$ and $\phi$ vs $f$ for $k^{*}=1.05$, respectively.
Circles  (stars) correspond the increase 
(decrease) of $f$
starting from an(a) ordered (disordered) phase.  The same
in panels
$(c)$ and $(d)$, but for $k^{*}=1.1$. Insets: Top and bottom
show $|m|$ vs $f$ for 
$k^{*}=1.3$ and the probability distribution $P(m)$ 
vs $m$ for distinct $f$'s and $N=3600$, respectively.}
\label{fig3}
\end{figure}

As  in the MFT, 
numerical simulations  also exhibit an additional phase transition  for
large $\theta$ and intermediate sets of $k^*$ (exemplified 
in Fig. \ref{fig3} for $\theta=0.45$).
Our results for low $f$ show a hysteretic loop signaling 
a phase coexistence between two synchronized phases
(see panels $(a)$ and $(c)$ for $k^{*}=1.05$
and $1.1$, respectively),
whereas by further increasing  $f$, $|m|$
vanishes continuously.
To achieve complementary
information about the hysteretic loop, we evaluate the difference
of magnetization restricted to the
subsets of nodes with and without inertia ($m_\theta$ and $m_0$, respectively)
given by $\phi=m_{\theta}-m_{0}$. For sufficient low $f$,
$m_\theta \approx m_0 \approx 1$, consistent with a full ordered phase.
The jump of $\phi$ to a moderate value indicates
that the nodes with inertia become
unsynchronized, but the vertices absent of inertia
remains ordered. Thus, in similarity
with the MFT, we observe that the system, in fact, 
exhibits a {\em partial}  synchronization. 
Although for $k^{*}=1.05$ the decrease of 
$f$ does not lead the system to the full synchronization
 (in similarity with the original 
order-disorder transition for large $\theta$), a closed
 hysteretic loop is observed for $k^{*}=1.1$.  A bimodal distribution
 in the hysteretic region (inset) reinforces a discontinuous transition
 between partially-ordered and ordered phases. As expected for the ER,
 the additional transition is absent for $k^{*}=1.3$ (inset).

\begin{figure}[h!]
  \epsfig{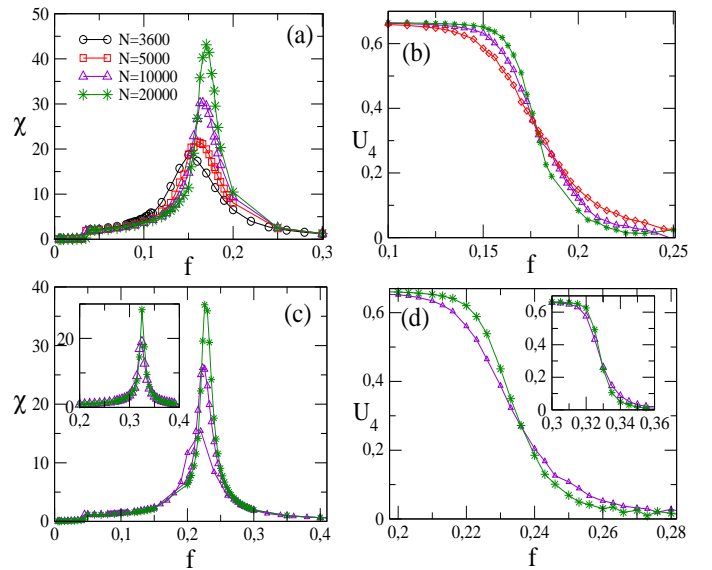}
\caption{For the ER network and $\theta=0.45$,
panels $(a)$ and $(b)$ show
$\chi$ and $U_4$ vs $f$ for $k^{*}=1.05$. 
Panels $(c)$ and $(d)$, the same but
for $k^{*}=1.1$. Insets: Results for $k^{*}=1.3$.}
\label{fig4}
\end{figure}

Now, we employ a finite size scaling analysis to characterize the order-disorder
phase transition.
In Fig. \ref{fig4}, we plot $\chi$ and $U_4$ for distinct network sizes $N$. We observe that they exhibit the typical
behaviors expected for continuous phase
transitions: the variance $\chi$ presents a maximum increasing with 
$N$ and their positions also systematically deviates on $N$.
Analysis of $U_4$  (Fig. \ref{fig4}: panels $(b)$, $(d)$ and its inset) show crossings
at $f_c=0.179(2)$, $f_c=0.236(3)$ and $f_c=0.330(5)$ with $U_4=0.27(2)$,
close to the values found in Ref. \cite{pereira}. Note that all
 critical points can be clearly distinguished from their previous
hysteric loops.

\begin{figure}[h!]
\epsfig{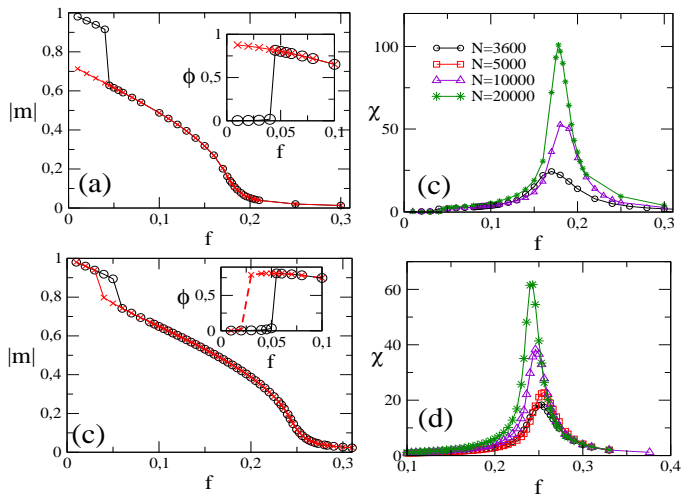}
\caption{For the PL network and $\theta=0.45$, 
$|m|$ versus  $f$ for $k^{*}=1.05$ [panel $(a)$]
  and $1.3$ [panel $(c)$].
Circles  (stars) correspond the increase 
(decrease) of $f$ starting from an(a) ordered (disordered) phase.
  Insets show the order parameter $\phi$ vs
$f$. Panels $(b)$ and $(d)$ shows the variance $\chi$
vs $f$ for distinct $N$'s.}
\label{fig5}
\end{figure}

Similar trends are visualized in Fig. \ref{fig5}  where we show
the results for  PL networks
with  $k^{*}=1.05$ and $1.3$, respectively. However, in this case,
the existence of hubs prolongs the partially synchronized
discontinuous transition for a larger set of restrictions than
the observed in ER networks.

\begin{figure}[h!]
\epsfig{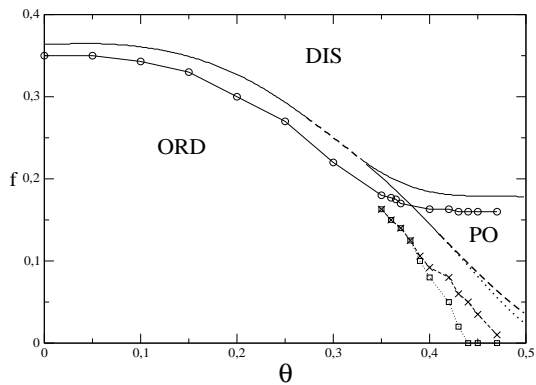}
\caption{Phase diagram $f \times \theta$ for partial
inertia $k^*=1.05$ and  ER network with $\langle k \rangle=20$. 
ORD, DIS and PO denote the ordered, disordered
and partially-ordered phases, respectively. Full lines  correspond
to the continuous  transitions, whereas dashed (dotted) 
lines correspond to the increase  (decrease) of $f$
starting from an(a) ordered (disordered) phase.}
\label{fig6-1}
\end{figure}

\begin{figure}[h!]
\epsfig{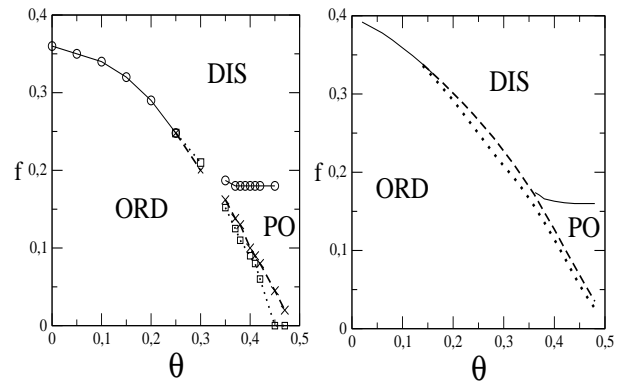}
\caption{Phase diagram $f \times \theta$ for partial
inertia $k^*=1.05$ and  PL network with $\langle k \rangle=20$. 
In the left and right panels, numerical simulations
for $N=20000$ and MFT, respectively.
ORD, DIS and PO denote the ordered, disordered
and partially-ordered phases, respectively.
Full lines  correspond
to the continuous  transitions, whereas dashed (dotted) 
lines  correspond to the increase  (decrease) of $f$
starting from an(a) ordered (disordered) phase.}
\label{fig7-1}
\end{figure}

In Fig. \ref{fig6-1} we plot the phase diagram for the ER and
$k^*=1.05$. As expected, both MFT and
numerical simulations predict continuous transition between
ordered (ORD) and disordered (DIS) phases  for small $\theta$.
MFT predicts the appearance of the partially ordered (PO)
for $\theta\ge\theta_c=0.330(1)$, whose transition is
 continuous in the interval 
$\theta_c \le \theta \le 0.41$
and  discontinuous for $\theta > 0.41$. Numerical
simulations  exhibit an additional peak of $\chi$ (absent of hysteresis),
consistent to the emergence of the PO for 
$0.36 \le \theta \le 0.39$ and a clear hysteretic loop for $\theta > 0.39$.
Despite the excellent qualitative agreement between approaches, 
it  is worth mentioning the difficulty of locating 
(and classifying) the PO phase transition for $\theta_c \le \theta \le 0.36$
under numerical simulations. 
Another point to mention concerns that the PO-DIS transition is 
always continuous and  practically independent on the inertia
for large $\theta$.  A qualitative similar phase diagram is
shown in Fig. \ref{fig7-1} for the PL case. However, the ordered-PO
transition line is always discontinuous, in qualitative agreement
with MFT predictions.

\section{Conclusions}

Recently, Chen et al. \cite{chen2} have found that inertia is responsible 
for the appearance of an abrupt transition
 in the majority vote model in complex networks.
In the present work, we advance by scrutinizing the inertia
acting only in the most connected nodes.
We show, through mean field analysis and numerical simulations
for homogeneous and heterogeneous networks,
that  $partial$ inertia can change
the system behavior depending on the inertia strength.

Our results also reveal that although  relevant  inertia rates 
are required for preserving the discontinuous transitions
for homogeneous networks, this is not the
case of heterogeneous structures, in which a rather small  fraction a
($17 \%$) already promotes an abrupt behavior. In other
words, by including only such above fraction in the sites
with larger degrees, the phase transition is discontinuous. This shares
some similarities
with the KM model, in which a positive frequency-degree
correlation included only in the hubs
is sufficient for sustaining an explosive synchronization
\cite{saa}.

A second remarkable effect of partial inertia concerns 
in the appearance of an additional phase characterized 
by a partial ordering of the system. The nature of the phase 
transition from the disordered phase to this partially ordered
 phase depends on the inertia strength. Therefore, there is a region 
in the phase diagram in which we observe two phase transitions: a 
continuous transition from the disordered to the partially ordered 
phase, and a discontinuous transition from the latter to the full 
ordered phase.

As pointed in \cite{chen2}, behavioral inertia is an essential
characteristic of human being and animal groups. Therefore,
inertia can be a  significant  ingredient in transitions
that arise in social systems \cite{social}, such as the
emergence of a common culture \cite{social2} or the appearance of
consensus \cite{sood} and decision-making systems \cite{couzin}. Our results suggest that inertia only in a small fraction of the population can 
produce dramatic effects if it is concentrated in the most connected individuals.

\end{document}